\documentclass[11pt,twoside]{article}
\usepackage{newpasp}
\usepackage{graphicx}
\def\edcomment#1{\iffalse\marginpar{\raggedright\sl#1\/}\else\relax\fi}
\reversemarginpar

\begin{document}
\title{Massive Molecular Outflows}
 \author{H. Beuther, P. Schilke, K.M. Menten}
\affil{Max-Planck-Institut f\"ur Radioastronomie, Auf dem H\"ugel 69, 53121 Bonn, Germany}
\author{C.M. Walmsley}
\affil{Osservatorio Astrofisico di Arcetri, Largo E. Fermi, 50125 Firenze, Italy}
\author{T.K. Sridharan}
\affil{Harvard-Smithsonian Center for Astrophysics, 60 Garden Street, MS 78, Cambridge, MA 02138, USA}
\author{F. Wyrowski}
\affil{Department of Astronomy, University of Maryland, College Park, USA}

\begin{abstract}
Well established correlations of low-mass flows show a continuity in
the high-mass regime, and mean accretion rates for
$L>1000~\rm{L}_{\odot}$ are $\dot{M}_{\rm{accr}}\sim
10^{-4}~\rm{M_{\odot}~yr}^{-1}$. We also present a tight correlation
between outflow masses and core masses, which implies that the
accretion rate is to first order a linear function of
the core mass. All those results support the idea that similar
physical processes drive outflows of all masses.
\end{abstract}

\vspace{-1cm}
\section{Introduction}
With the aim to understand the role of massive outflows in high-mass
star formation, we mapped 26 massive star formation regions in
CO(2--1) with a spatial resolution of $11''$. A detailed analysis of
the data is presented by Beuther et al. (subm. to A\&A). Here we are
just outlining the major statistical results of the study.

\section{Main results}

In 21 of the observed sources show bipolar structure, they are very
massive and energetic and show higher collimation factors than
previous studies. Mean outflow parameters are:

\begin{tabular}{lr|lr}
mass             & 50 [M$_{\odot}$] &  $\dot{M}_{\rm{out}}$ & $7.5\times 10^{-4}$ [M$_{\odot}$ yr$^{-1}$] \\
momentum    & 2200 [M$_{\odot}$ km s$^{-1}$]& mech. force & 0.04 [M$_{\odot}$ km$^2$ s$^{-2}$ yr$^{-1}$]\\
energy  & 6000 [M$_{\odot}$ km$^2$ s$^{-2}$] & col. factor & 2.1  \\
time                             & 70000 [yr]\\
\end{tabular}

\noindent Fig. 1 presents comparisons of different quantities within our sample
to previous studies in the low- and high-mass regime. Fig. 1(a) shows
that previously determined correlations in the low-mass regime between
the mechanical force $F_{\rm{CO}}$ and the core mass $M_{\rm{core}}$
seem to be valid for massive star formation regions as well. We
consider this as support for similar physical processes driving
outflows of all masses.

\begin{figure}[ht]
\includegraphics[bb=282 18 527 807,angle=-90,width=13.3cm]{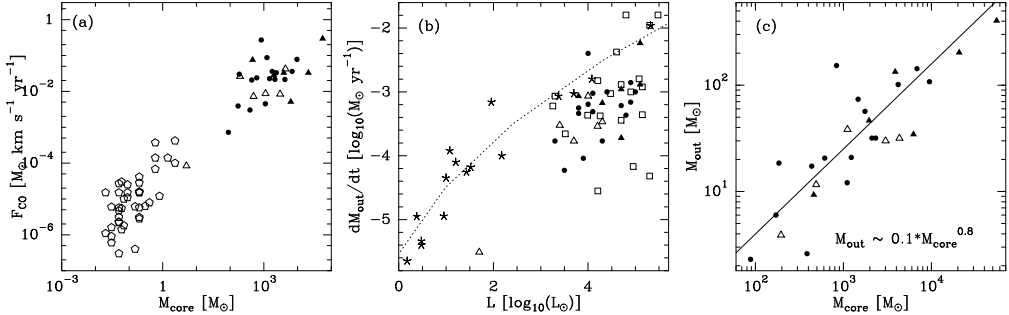}
\caption{{\bf (a)} mechanical force versus core mass, {\bf (b)} mass entrainment rate versus luminosity and {\bf (c)} outflow mass versus core mass. The circles show the sources with known distance, and the full and open triangle represent sources with far and near distance, respectively. The pentagons show data from Bontemps et al. (1996), the asterisks data from Cabrit \& Bertout (1992), and the squares are compiled by Churchwell (2000). The dashed line in (b) is the fit taken from Shepherd \& Churchwell (1996).} 
\end{figure}

It looks different in Fig 1(b), where we plot the mass entrainment
rate versus the luminosity. Obviously, at the high-mass end those data
are not correlated. While outflows in low-mass regions can be
identified usually with one protostar, massive stars form in clusters,
and the resolution is not sufficient to resolve the driving sources
exactly. Thus, possibly the flows are driven from one of the youngest
sources whereas the main part of luminosity might stem from a more
evolved source, which could already have reached the main
sequence. But so far this is speculation and has to be investigated at
higher resolution in detail. Our data indicate high mean accretion
rates $\dot{M}_{\rm{accr}}\sim 10^{-4}~\rm{M_{\odot}~yr}^{-1}$ for $L>
1000~\rm{L}_{\odot}$ in a few cases larger than
$10^{-3}~\rm{M_{\odot}~yr}^{-1}$.

Furthermore, we see a tight correlation between the mass of the
molecular outflow and the core mass (Fig. 1(c)). The ratio of both
quantities is rather constant with a mean of 0.04. As outlined in
Beuther et al. (submitted to A\&A) this correlation implies that the
product of the accretion efficiency
$f_{\rm{acc}}=\dot{M}_{\rm{acc}}/(M_{\rm{core}}/t_{\rm{ff}})$ and the
deflection efficiency
$f_{\rm{r}}=\dot{M}_{\rm{jet}}/\dot{M}_{\rm{acc}}$ is about constant
for all star forming regions, and the accretion rate is
an approximately linear function of the core mass:
${\dot{M}_{\rm{acc}}= \frac{f_{\rm{acc}}}{t_{\rm{ff}}} M_{\rm{core}}
= 7 \cdot 10^{-8} \times M_{\rm{core}}\ [\rm{M}_{\odot}
~\rm{yr}^{-1}]}.$

\end{document}